# Chaotic Motion of Ions in Finite-amplitude Low-frequency Alfvén Waves

Jingyu Peng[1], Jiansen He[1,2], and Rong Lin[1]
[1] School of Earth and Space Sciences, Peking University, Beijing, People's Republic of China; jshept@pku.edu.cn
[2] State Key Laboratory of Solar Activity and Space Weather, National Space Science Center, Chinese Academy of Sciences, Beijing, People's Republic of China
*Received 2025 November 20; revised 2026 January 2; accepted 2026 January 15; published 2026 February 20*

## Abstract

Finite-amplitude low-frequency Alfvén waves (AWs) are ubiquitous in space plasmas, where they play a key role in the transport and dissipation of energy, particularly in the heating of ions in the solar corona and solar wind. In this study, we investigate the nonlinear interaction between ions and obliquely propagating AWs. When the wave amplitude and propagation angle lie within specific ranges, ion motion becomes chaotic. We quantify this behavior using the maximum Lyapunov exponent ($\lambda_m$) and define a new parameter, the chaos ratio (CR), to describe the fraction of chaotic particles across different initial states. The global chaos threshold is determined as the contour CR = 0.01. Analysis of magnetic moment variations reveals that the physical origin of chaos is pitch-angle scattering induced by *wave-driven field-line curvature* (WFLC), which disrupts adiabatic invariance and leads to stochastic ion energization. The onset condition for chaos can be expressed by an effective relative curvature radius, $P_{\text{eff.}} < 25$. This analytical criterion delineates the boundary of the chaotic region in the ($k_x$, $k_z$, $B_w$) parameter space and agrees well with numerical results. The identified WFLC mechanism provides a new physical pathway for converting macroscale Alfvénic disturbances into microscopic ion heating. This analysis offers a simplified model that illustrates a plausible ion energization mechanism in Alfvénic turbulent plasmas, including those associated with solar wind switchbacks and coronal fluctuations. These results highlight a universal chaotic process that may underlie stochastic heating in heliospheric and astrophysical plasmas.

## 1. Introduction

The heating of ions in the corona and solar wind is a significant topic in heliospheric physics. Cyclotron resonance between ions and Alfvén waves (AWs) with frequencies near the ion gyrofrequency is considered a key mechanism for ion heating (S. R. Cranmer 2001; S. P. Gary et al. 2001, 2006; C. Y. Tu & E. Marsch 2001; J. V. Hollweg & P. A. Isenberg 2002). However, the precise contribution of cyclotron resonance to ion heating remains uncertain. There is no direct evidence that such high-frequency fluctuations in the corona and solar wind possess sufficient energy to heat ions. Observations indicate that most of the AW energy in the corona is concentrated in the low-frequency band (I. V. Chashei et al. 2000; B. De Pontieu et al. 2007; S. Tomczyk et al. 2007; D. B. Jess et al. 2009), and the energy of solar wind AW turbulence is also primarily located at large scales (E. J. Smith et al. 1995; C. Y. Tu & E. Marsch 1995; S. D. Bale et al. 2005; R. Bruno & V. Carbone 2013). Moreover, due to the perpendicular cascade of solar wind turbulence, the transfer of wave energy to higher frequencies is highly inefficient (S. R. Cranmer & A. A. van Ballegooijen 2003; G. G. Howes et al. 2008; F. Sahraoui et al. 2009; B. D. G. Chandran et al. 2010; J. He et al. 2013).

Although low-frequency AWs are ineffective at heating ions through cyclotron resonance, they can energize ions via the stochastic heating mechanism. When the wave amplitude exceeds a certain threshold, ion motion becomes chaotic (L. Chen et al. 2001). This mechanism constitutes an important component of the kinetic Alfvén wave (KAW) ion heating model (M. K. Chettri et al. 2024). Ion heating by large-amplitude, low-frequency AWs has also been supported by in situ solar wind observations (Y. J. Rivera et al. 2024). These waves are believed to originate from the Sun or its vicinity (B. D. G. Chandran et al. 2010), suggesting that the stochastic heating mechanism can naturally operate in both the corona and the solar wind. Furthermore, observations indicate that proton heating occurs within solar wind switchbacks (Q. Luo et al. 2023), which behave as large-amplitude Alfvénic disturbances (L. Matteini et al. 2014, 2015). Stochastic ion heating by low-frequency monochromatic AWs with different polarization states has been investigated in several studies (L. Chen et al. 2001; O. Y. Kolesnychenko et al. 2005; X. Lv et al. 2007; J. Sun et al. 2014), where the Poincaré surface of section (PSOS) is used to distinguish between regular and chaotic trajectories in phase space. However, PSOS analysis becomes inadequate in the presence of multiple wave modes (Q. Lu & L. Chen 2009), motivating the introduction of quantitative indices for characterizing chaos. Z. Guo et al. (2008) applied Lie perturbation theory to examine higher-order subgyrofrequency resonances. According to the Chirikov criterion (B. V. Chirikov 1960; S. M. Soskin et al. 2003), global chaos emerges when adjacent resonances overlap, leading to stochastic ion heating. At sufficiently large amplitudes, ions can also be trapped within magnetic-mirror-like field structures formed by low-frequency AWs (C.-B. Wang et al. 2013).

Previous studies have described chaotic motion as a result of subgyrofrequency resonance (L. Chen et al. 2001; R. White et al. 2002; Z. Guo et al. 2008), but they have not connected chaotic motion and changes in magnetic moment to the effect of field-line curvature (FLC). Establishing this link is critical for advancing our understanding of chaotic dynamics in low-frequency waves. Additionally, previous studies adopted wave amplitude exceeding a certain threshold as the condition for the onset of chaos (L. Chen et al. 2001; R. White et al. 2002; O. Y. Kolesnychenko et al. 2005; X. Lv et al. 2007; Q. Lu & L. Chen 2009; C.-B. Wang et al. 2013; K. Fan et al. 2021). However, this threshold varies depending on the wave parameters, highlighting the need for a more universal condition. This work aims to address these unresolved challenges. In this study, ions are assumed to exert negligible influence on the waves and are therefore treated as test particles—for example, minor ions in the solar wind. We perform test-particle simulations to explore the chaotic motion of ions in obliquely propagating, monochromatic, finite-amplitude, low-frequency AWs. Our objective is to quantitatively evaluate how wave parameters control field-line deformation or wave-induced field-line curvature (WFLC) and the resulting particle chaos.

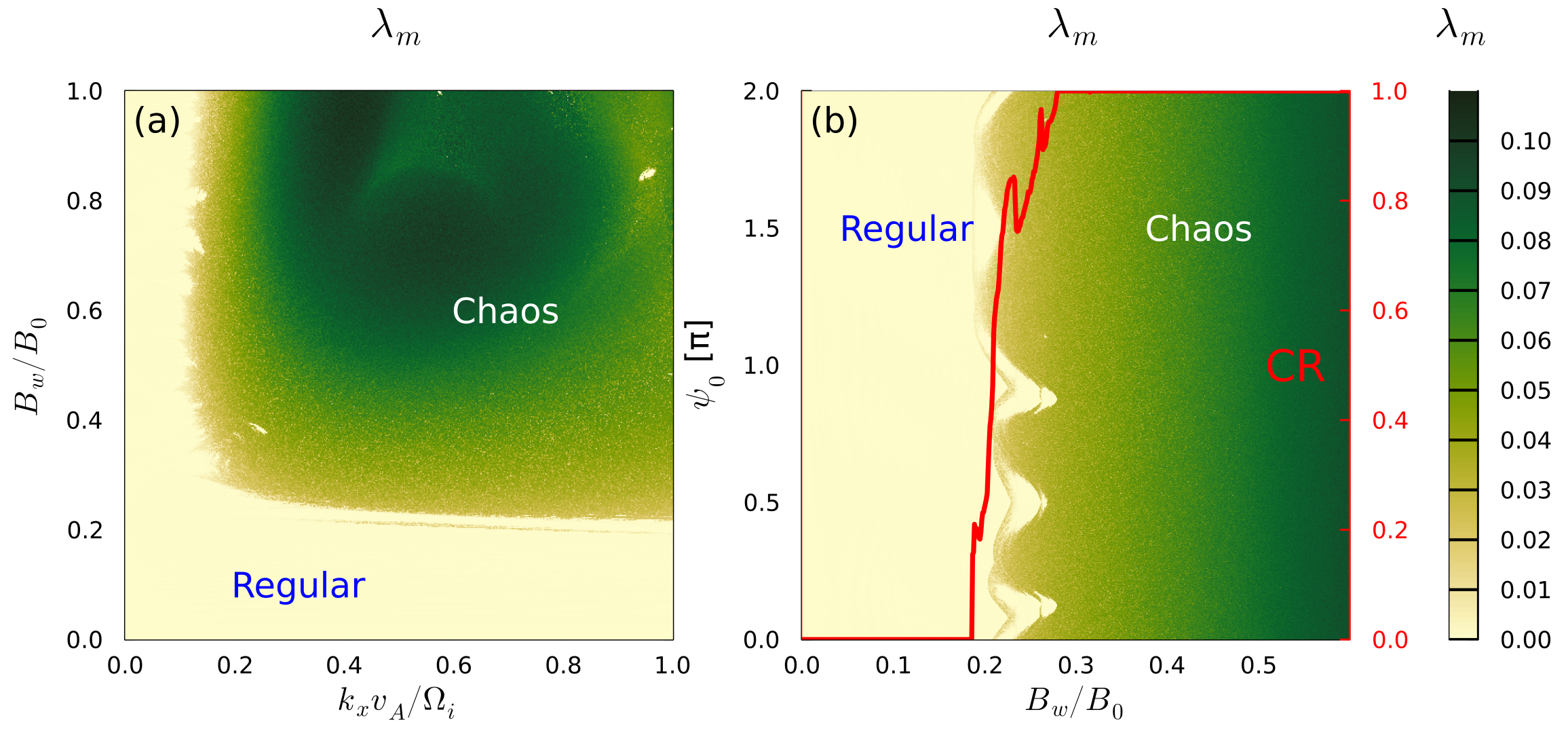

**Figure 1.** Chaos quantification indices. (a) $\lambda_m$ in the $k_x^* - B_w^*$ parameter space, with $k_z^* = 0.25$. The initial state is (0, 0, 0, −1). (b) $\lambda_m$ for different values of $B_w^*$ and initial phases $\psi_0$, with the particle's initial velocity set to $(0, 0, -v_A)$, $k_x^* = 0.5$, and $k_z^* = 0.25$. The red line is the chaos ratio (CR) among all 1000 initial states, where particle motion is classified as chaotic when $\lambda_m > 0.0025$.

## 2. Method

A left-hand circularly polarized AW is considered, with the propagation angle between the wavevector $\boldsymbol{k} = (k_x,\ 0,\ k_z)$ and the constant background magnetic field $\boldsymbol{B_0} = B_0\hat{\boldsymbol{z}}$ is $\alpha$. The wave dispersion relation in the plasma frame is $\omega = k_z v_A$, where $\omega$ is the wave frequency, $v_A = \frac{B_0}{\sqrt{\mu_0\rho_m}}$ is the Alfvén speed, $\mu_0$ is the vacuum magnetic permeability, and $\rho_m$ is the plasma mass density. In the remainder of this paper, our analysis is conducted in the wave frame, which is assumed to be moving at $v_A\hat{\boldsymbol{z}}$ relative to the plasma frame. The advantage of the wave frame is that the wave magnetic field is time-independent, and the wave electric field vanishes under the low-frequency AW conditions, simplifying the analysis (L. Chen et al. 2001). The wave magnetic field (L. Chen et al. 2001; Q. Lu & L. Chen 2009):

$$\boldsymbol{B_w} = B_w(-\cos\alpha\sin\psi\hat{\boldsymbol{x}} + \cos\psi\hat{\boldsymbol{y}} + \sin\alpha\sin\psi\hat{\boldsymbol{z}}), \tag{1}$$

where the phase $\psi = k_x x + k_z z$ and $B_w$ is the wave amplitude. The magnetic field experienced by a particle at position $\boldsymbol{x} = (x,\ y,\ z)$ is determined by $\psi(\boldsymbol{x})$. The governing equation of motion for an ion of species $i$ is

$$\dot{\psi} = k_x v_x + k_z v_z, \tag{2}$$

$$\dot{\boldsymbol{v}} = \Omega_i \boldsymbol{v} \times (\hat{\boldsymbol{z}} + \boldsymbol{B_w}/B_0), \tag{3}$$

$$\dot{\boldsymbol{x}} = \boldsymbol{v}, \tag{4}$$

where the gyrofrequency $\Omega_i = q_i B_0/m_i$, with $q_i$ and $m_i$ being the charge and mass of species $i$, respectively. Equations (2) and (3) form a complete ordinary differential equation (ODE) system describing the motion of ions in a four-dimensional dimensionless state space $\boldsymbol{s} = (\psi,\ v_x/v_A,\ v_y/v_A,\ v_z/v_A)$. This ODE system contains three dimensionless parameters: $k_x^* = k_x v_A/\Omega_i$, $k_z^* = k_z v_A/\Omega_i$, and $B_w^* = B_w/B_0$. According to Equation (3), $dv^2/dt = 0$. Thus, the ion speed $v$ is constant in the wave frame, and the particle trajectories in velocity space lie on a spherical shell with radius $v$.

## 3. Results

An important feature of chaos is the sensitive dependence of trajectories on initial states. The distance in state space between two initially close trajectories $|\boldsymbol{\delta}|(t) = |\boldsymbol{s_2}(t) - \boldsymbol{s_1}(t)|$ grows exponentially over time $|\boldsymbol{\delta}|(t) \sim \exp(\lambda_m t)$, where $\lambda_m$ is the maximum Lyapunov exponent (G. Benettin et al. 1980; K. Geist et al. 1990; A. M. Lyapunov 1992; G. Datseris 2018; G. Datseris & U. Parlitz 2022). In contrast, for regular motion, $|\boldsymbol{\delta}|$ remains small and $\lambda_m \approx 0$. Thus, $\lambda_m$ can be used to quantify the degree of chaos. The method we use to calculate $\lambda_m$ is the "H2" method described in K. Geist et al. (1990). We consider particles starting from the initial state (0, 0, 0, −1). For $k_z^* = 0.25$, Figure 1(a) shows $\lambda_m$ in the parameter space $B_w^* - k_x^*$. When $k_x^*$ and $B_w^*$ are large, $\lambda_m$ has a significant positive value, indicating chaotic particle motion.

We study the dependence of chaos on particle states. A total of 1000 particles initially at rest in the plasma frame are considered, so their initial velocity is $\boldsymbol{v} = (0, 0, -v_A)$, and their initial phases $\psi_0$ are uniformly distributed in $[0, 2\pi]$. Figure 1(b) shows $\lambda_m$ of particles with different parameters $B_w^*$ and initial phases $\psi_0$. We consider particle motion to be chaotic when $\lambda_m > 0.0025$. This threshold is established based on the distribution of $\lambda_m$ values observed in simulations of both regular and chaotic motion, ensuring a clear distinction between the two types of motion. Using this criterion, we can calculate the ratio of particles exhibiting chaotic motion among those with different initial states, which we refer to as the chaos ratio (CR),

$$\mathrm{CR} = \frac{1}{N}\sum_{i=1}^{N} \mathbb{I}_{\lambda_m^i > 0.0025}, \tag{5}$$

where $\lambda_m^i$ is the maximum Lyapunov exponent of the $i$th particle, $N$ is the total number of particles with different initial states, and $\mathbb{I}_{\lambda_m^i > 0.0025}$ is the indicator function,

$$\mathbb{I}_{\lambda_m^i > 0.0025} = \begin{cases} 1 & if \ \ \lambda_m^i > 0.0025, \\ 0 & if \ \ \lambda_m^i \leqslant 0.0025. \end{cases}$$

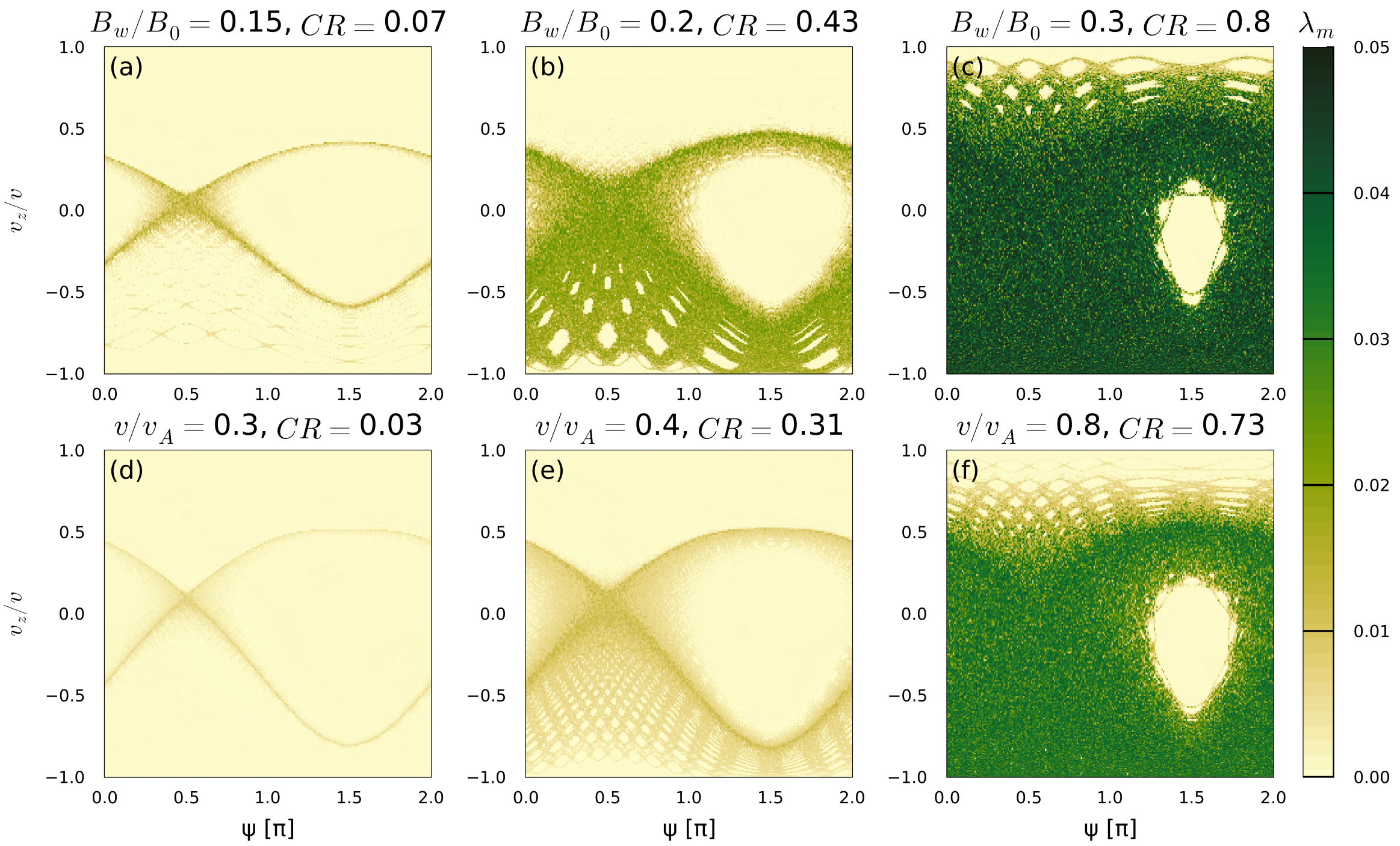

**Figure 2.** $\lambda_m$ in the state space ($\psi$, $v_z/v$). (a)–(c) $B_w^* = 0.15, 0.2, 0.3$, parameters $k_x^* = 0.5$, $k_z^* = 0.25$, ion speed $v/v_A = 1$. (d)–(f) ion speed $v/v_A = 0.3, 0.4, 0.8$, parameters $B_w^* = 0.3$, $k_x^* = 0.5$, $k_z^* = 0.25$. CR among particles with different initial $\psi$ and $v_z/v$ are indicated.

For smaller values of $B_w^*$, all particles' motion are regular, $\mathrm{CR} = 0$; for larger values of $B_w^*$, all particles' motion are chaotic, $\mathrm{CR} = 1$; there exists an interval of $B_w^*$ (specifically, $[0.19, 0.28]$ in Figure 1(b)) where the regularity or chaos depends on $\psi_0$, and only a portion of the particles exhibit chaos behavior, with $0 < \mathrm{CR} < 1$. Previous studies on chaos through PSOS rely on manual judgment to identify chaos behavior, which introduces a degree of uncertainty (L. Chen et al. 2001; Q. Lu & L. Chen 2009). By quantifying chaos through the metrics $\lambda_m$ and CR, we achieve more accurate and efficient identification of chaos while effectively delineating chaos regions in both the parameter space ($k_x^*$, $k_z^*$, $B_w^*$) and the state space ($\psi$, $v_x/v_A$, $v_y/v_A$, $v_z/v_A$).

We also investigate the chaotic behavior of particles with different initial phases and velocities. As shown in Figures 2(a) and (d), when the wave amplitude $B_w/B_0$ or ion speed $v$ is small, chaos is confined to thin layers associated with nonlinear higher-order cyclotron resonances at subgyrofrequencies (L. Chen et al. 2001; R. White et al. 2002; Z. Guo et al. 2008). As $B_w/B_0$ or $v$ increases, these chaotic layers expand and begin to overlap with neighboring layers, forming broader chaotic regions, as shown in Figures 2(b), (c), (e), and (f). Concurrently, the CR increases as these chaotic regions grow in the state space. The calculated results of $\lambda_m$ closely match the chaotic regions identified through PSOS analysis (R. White et al. 2002; X. Lv et al. 2007; Z. Guo et al. 2008), further confirming the effectiveness of $\lambda_m$ in detecting chaos.

The chaotic motion of a particle is studied. In Figure 3, the particle undergoes spiral motion when $t < 1500/\Omega_i$. The parallel velocity $v_\parallel = \boldsymbol{v} \cdot \frac{\langle \boldsymbol{B} \rangle_g}{|\langle \boldsymbol{B} \rangle_g|}$ exhibits discontinuities (see Figure 3(a)), where $\langle \cdot \rangle_g(t) = \frac{1}{T_g}\int_{t-T_g/2}^{t+T_g/2} \cdot dt'$ denotes an average over one gyroperiod $T_g = 2\pi/\Omega_i$, and $\boldsymbol{B} = \boldsymbol{B_0} + \boldsymbol{B_w}$ is the total magnetic field. This discontinuity can also be interpreted as pitch-angle scattering, since the pitch angle $\delta = \arccos(v_\parallel/v)$ also undergoes abrupt changes. The abrupt change of $v_\parallel$ occurs near the field minimum (see Figure 3(b)). We calculated the curvature radius of the field lines

$$R_c(\psi;\ B_w,\ \boldsymbol{k}) = \frac{|\boldsymbol{B}|^2}{|\boldsymbol{B} \cdot \nabla \boldsymbol{B}|} = \frac{|\boldsymbol{B}|^2}{k_z B_w B_0} = \frac{1 + B_w^{*2} + 2B_w^* \sin\alpha \sin\psi}{k_z B_w^*}. \tag{6}$$

Thus, the position of the field minimum also corresponds to the position of the minimum $R_c$, which represents the location where the magnetic field changes most rapidly in spatial space. The minimum $R_c$ occurs at the phase $\psi_m = \frac{3}{2}\pi + 2n\pi,\ n \in Z,$

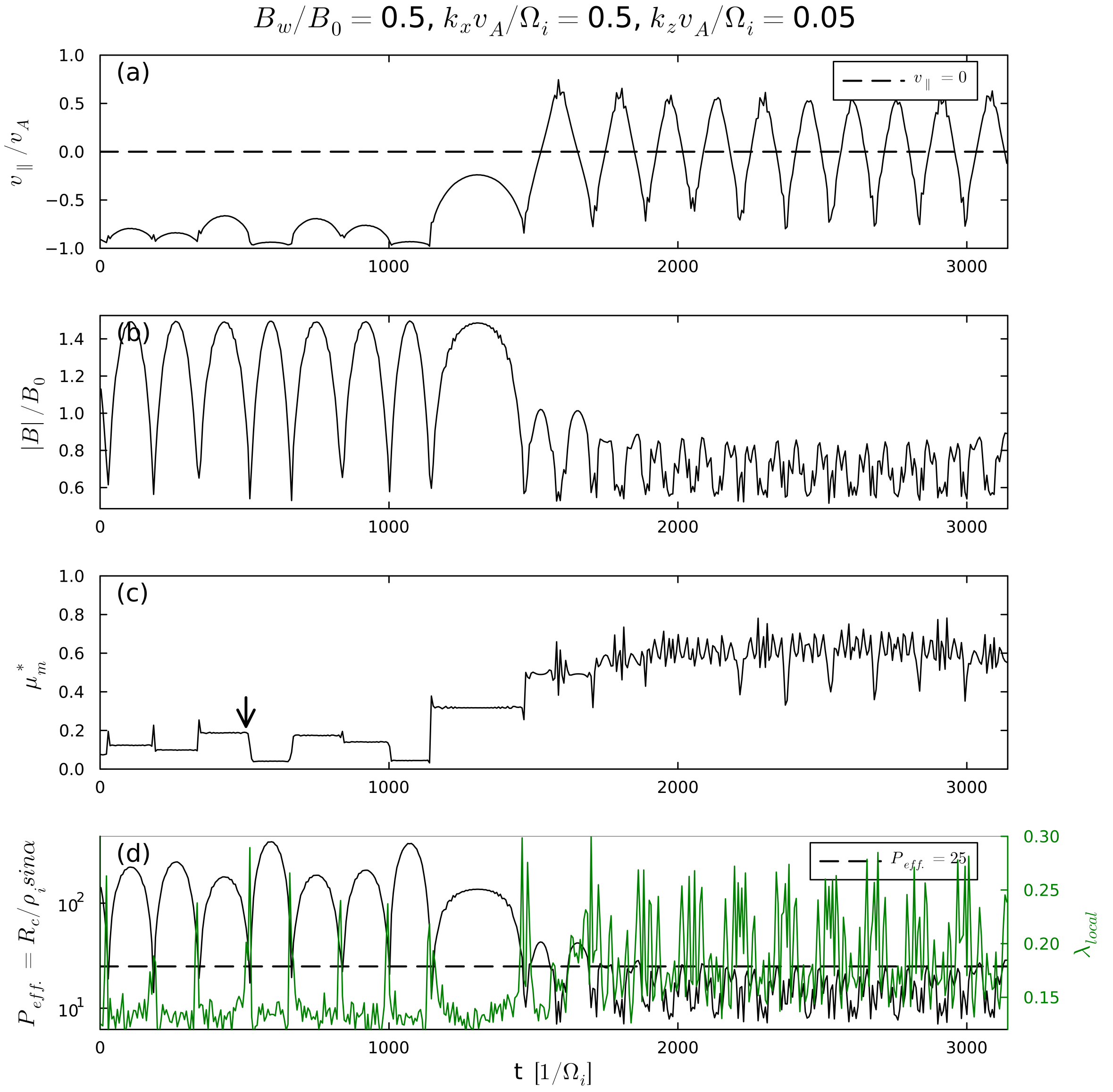

**Figure 3.** Time series of chaotic motion with parameters $B_w^* = 0.5$, $k_x^* = 0.5$, $k_z^* = 0.05$. The initial state of the particle is (0, 0, 0, −1). (a) Parallel velocity $v_\parallel$, the dashed line indicates the positions of velocity reversal, i.e., $v_\parallel = 0$. (b) Magnitude of the magnetic field $|\boldsymbol{B}|/B_0$. (c) $\mu_m^*$, the arrow marks the position of the $\mu_m^*$ change plotted in Figure 4(b). (d) Black line: effective relative curvature radius $P_{\rm eff.}$, the dashed line indicates $P_{\rm eff.} = 25$. Green line: the maximum local Lyapunov exponent $\lambda_{\rm local}$ during one gyroperiod.

and is given by

$$R_c^m(B_w,\ \boldsymbol{k}) = \frac{1}{k_z}\left(B_w^* + \frac{1}{B_w^*} - 2\sin\alpha\right). \tag{7}$$

In addition to spiral motion, the particle can enter bounce motion orbits. In Figure 3(a), when $t > 1500/\Omega_i$, the parallel velocity passes through zero, indicating the mirror points of the bounce motion.

The magnetic moment $\mu_m$ is the adiabatic invariant of a charged particle moving in a magnetic field (J. D. Jackson 1998). Figure 3(c) shows the changes in the magnetic moment during the particle's motion. The dimensionless magnetic moment $\mu_m^*$ is calculated by

$$\mu_m^* = \frac{\frac{1}{2}\langle|\boldsymbol{v}_\perp|^2\rangle_g/v_A^2}{|\langle\boldsymbol{B}\rangle_g|/B_0}, \tag{8}$$

where the perpendicular velocity $\boldsymbol{v}_\perp = \boldsymbol{v} - v_\parallel\frac{\langle\boldsymbol{B}\rangle_g}{|\langle\boldsymbol{B}\rangle_g|}$. $\mu_m^*$ changes near the field minimum, as shown in Figures 3(b) and (c). When the magnetic field experienced by a particle changes significantly during a gyroperiod—that is, when the spatial scale of the magnetic field variation becomes comparable to the gyroradius—the conservation of the adiabatic invariant $\mu_m^*$ breaks down. Therefore, the condition for the breakdown of the conservation of $\mu_m^*$ can be expressed as

$$P_{\rm eff.} = \frac{||\nabla\boldsymbol{B}||_F}{||\nabla_\perp\boldsymbol{B}||_F}R_c/\rho_i = \frac{R_c/\sin\alpha}{\rho_i} < C, \tag{9}$$

where $C$ is a constant, $\rho_i = m_i v_\perp/q_i|\boldsymbol{B}| = v_\perp B_0/\Omega_i|\boldsymbol{B}|$ is the gyroradius, $||\cdot||_F$ is the Frobenius norm, and $\frac{||\nabla\boldsymbol{B}||_F}{||\nabla_\perp\boldsymbol{B}||_F} = \frac{1}{\sin\alpha}$ is the projection factor that maps the curvature radius onto the perpendicular direction, since the gyromotion is primarily in the perpendicular direction, $\nabla_\perp = (\mathbf{I} - \hat{e}_\parallel\hat{e}_\parallel^{\rm T})\nabla$ is the perpendicular nabla operator, $\hat{e}_\parallel$ is the unit vector in the background magnetic field direction. We call $P_{\text{eff.}}(\psi,\ \boldsymbol{v};\ B_w,\ \boldsymbol{k})$ the effective relative curvature radius.

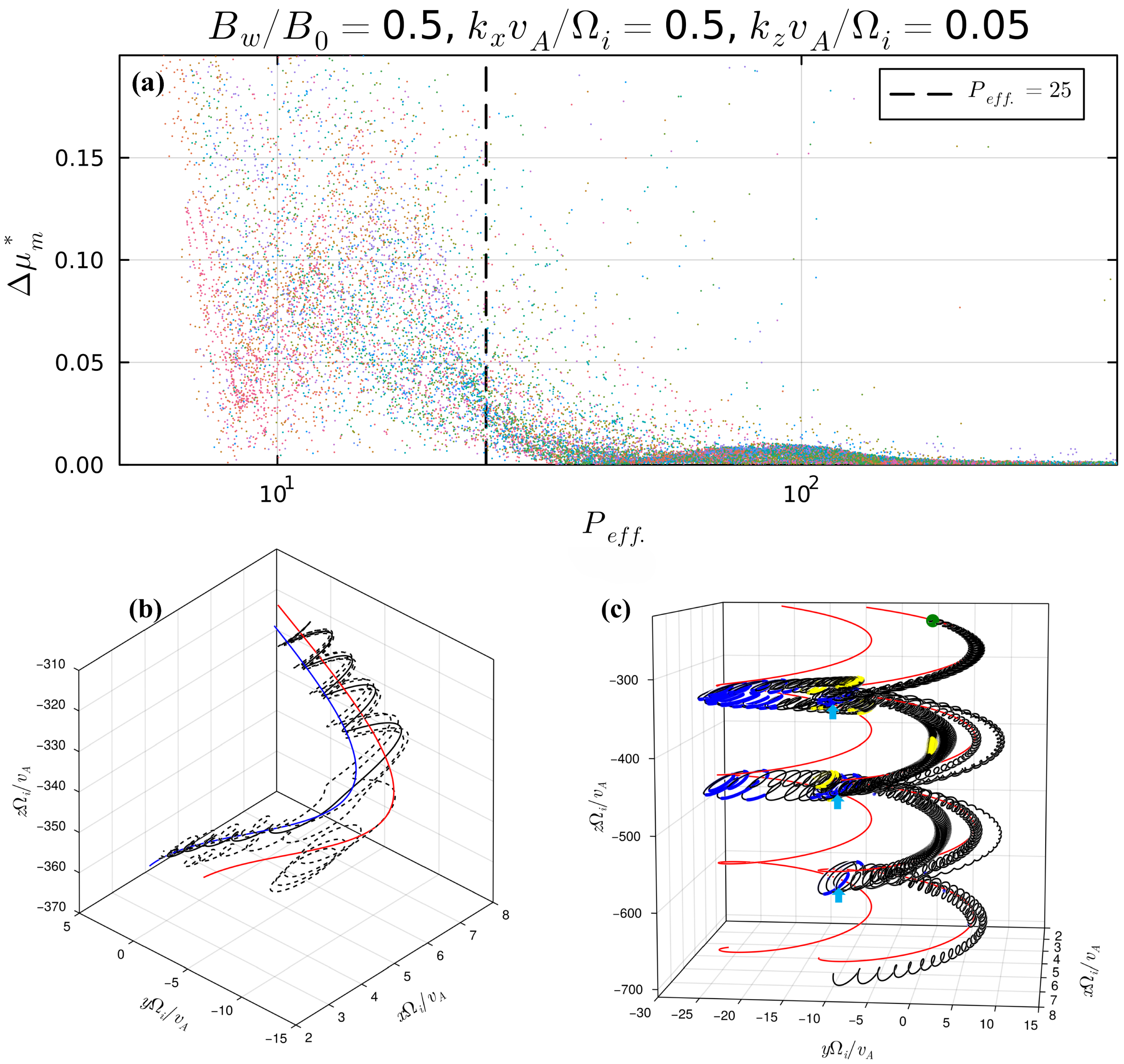

**Figure 4.** (a) Changes in the magnetic moment $\Delta\mu_m^*$ at different values of $P_{\text{eff.}}$, with different colors representing particles with different initial states. A total of 50 particles are considered, each with speed $v = v_A$, initial pitch angles $\theta_0$ uniformly distributed in $[0, \pi]$, initial azimuth angle $\phi_0 = 0$, and initial phases $\psi_0$ uniformly distributed in $[0, 2\pi]$. The dashed line indicates $P_{\text{eff.}} = 25$. Parameters are the same as Figure 3. (b) The particle's trajectory (black solid line) and magnetic field lines at the time corresponding to the arrow in Figure 3(c). The red line represents the field-line particle gyrating around before the $\mu_m^*$ change, while the blue line represents the field-line particle gyrating around after the $\mu_m^*$ change. The four black dashed lines indicate trajectories with slight differences in initial velocities. (c) Trajectories of nine initially adjacent particles over 100 gyroperiods (black lines). The particles start from the positions marked by the green dot. Two magnetic field lines are plotted for reference (red lines). Positions where $P_{\text{eff.}} < 25$ are marked in blue; pitch-angle scattering events that cause particle trajectory separation are marked with three light blue arrows; and positions where $|v_\parallel| < 0.05$ are marked in yellow, indicating the mirror points of the bouncing motion.

The change in $\mu_m^*$ between neighboring gyroperiods $\Delta\mu_m^* = \frac{1}{2}(|\mu_m^* - \mu_{m,-1}^*| + |\mu_m^* - \mu_{m,+1}^*|)$ is calculated, where $\mu_{m,-1}^*$ and $\mu_{m,+1}^*$ represent the values of $\mu_m^*$ of the previous and next gyroperiods, respectively. As shown in Figures 3(d) and 4(a), regardless of the particles' initial states, $\mu_m^*$ undergoes a significant change when $P_{\text{eff.}} < C$, $C \approx 25$. Figure 4(b) shows the particle's trajectory when $\mu_m^*$ changes (black solid line), corresponding to the time marked by the arrow in Figure 3(c). As the particle approaches the position of minimum $P_{\text{eff.}}$, see Figure 3(d), it ceases to gyrate around the red field line and enters a "transfer orbit" until it is captured by another field line (the blue line in Figure 4(b)). During this process, the quasiperiodic gyromotion is disrupted, and the adiabatic condition for the conservation of $\mu_m^*$ cannot be ensured. After the particles are recaptured, the pitch angle on the new field-line orbit may differ significantly from that on the previous orbit, resulting in a change in $\mu_m^*$. During this process, the particle's trajectory is highly sensitive to its initial state. If there are slight differences in the initial state, the particle may follow entirely different "transfer orbits" and new field-line orbits, as shown in Figure 4(b). This indicates that changes in $\mu_m^*$ exhibit chaotic behavior, which is also supported by the high values of the maximum local Lyapunov exponent $\lambda_{\text{local}}$ (H. D. I. Abarbanel et al. 1991) when $\mu_m^*$ changes, see Figure 3(d). Thus, the physical image of the chaos caused by low-frequency AWs is pitch-angle scattering caused by WFLC. We cannot accurately predict when a particle will cease to gyrate around the field line, nor can we accurately

determine which field line will recapture it. This uncertainty renders changes in $\mu_m$ random, explaining why chaotic motion is unpredictable over the long term. Additionally, we plot the trajectories of several initially adjacent particles over a longer time period, as shown in Figure 4(c). It is evident that these trajectories diverge significantly after multiple pitch-angle scattering events, which are marked with light blue arrows. Notably, some particles enter bounce motion orbits in Figure 4(c), causing their motion along the parallel direction to reverse; the mirror points are marked in yellow.

Chaos regions in the parameter space can be described by Equation (9). For ions with $v = v_{\mathrm{A}}$, we consider the minimum effective relative curvature radius $P^m_{\mathrm{eff.}} = R_c^m/(\sin\alpha\rho_i^m)$, where the maximum gyroradius $\rho_i^m = v_A/\Omega_i\sqrt{1 + B_w^{*2} - 2B_w^*\sin\alpha}$. The condition for the onset of global chaos is

$$P^m_{\mathrm{eff.}}(B_w^*, \boldsymbol{k}^*) = \frac{1}{k_z^* B_w^* \sin\alpha}(1 + B_w^{*2} - 2B_w^*\sin\alpha)^{3/2} < C. \tag{10}$$

If $k_x \ll k_z$, then $\sin\alpha \ll 1$, $P^m_{\mathrm{eff.}} \approx \frac{1}{k_x^*}\left(\frac{1+B_w^{*2}}{B_w^{*2/3}}\right)^{3/2}$ takes the minimum value when $B_w^* = \sqrt{2}/2$. Thus, the condition for Equation (10) to have solutions is $k_x^* > \frac{3\sqrt{3}}{2C}$. This implies that when the wave propagates quasi-parallelly, i.e., $k_x^* \approx 0$, chaos does not occur. On the other hand, if $k_x \gg k_z$, then $\sin\alpha \approx 1$, $P^m_{\mathrm{eff.}} \approx \frac{|1 - B_w^*|^3}{k_z^* B_w^*}$. The chaos threshold for $B_w^*$ is given by the smaller real root of $\frac{|1 - B_w^*|^3}{k_z^* B_w^*} = C$, $B_w^{*,-} = 1 - \left(\frac{k_z^* C}{2} + \sqrt{\frac{k_z^{*2}C^2}{4} + \frac{k_z^{*3}C^3}{27}}\right)^{1/3} - \left(\frac{k_z^* C}{2} - \sqrt{\frac{k_z^{*2}C^2}{4} + \frac{k_z^{*3}C^3}{27}}\right)^{1/3}$, which is independent of $k_x$. The solution for $P^m_{\mathrm{eff.}} = 25$ in $B_w^* - k_x^*$ parameter space is plotted in Figure 5.

Figure 5 shows CR in $k_x^* - B_w^*$ parameter space. Contour lines at $\mathrm{CR} = 0.01$ are drawn to delineate the region where particle motion begins to exhibit chaotic behavior. This result is very close to the stochastic heating threshold obtained by R. White et al. (2002) through PSOS. We find that this chaos region is also very close to the region where $P^m_{\mathrm{eff.}} < 25$, as shown by the red dashed lines. This confirms that chaotic motion arises from the disruption of gyromotion caused by WFLC, and $P^m_{\mathrm{eff.}} < C \approx 25$ indicates the presence of global chaos.

## 4. Discussion

We demonstrate that the physical image of ions' chaotic motion in finite-amplitude low-frequency AWs is pitch-angle scattering caused by WFLC, and chaos can be characterized by a single parameter $P_{\mathrm{eff.}}$. The regions of chaos in parameter space are analytically defined. Earlier studies proposed that chaos occurs when the wave amplitude exceeds a certain threshold (L. Chen et al. 2001; R. White et al. 2002; O. Y. Kolesnychenko et al. 2005; X. Lv et al. 2007; Q. Lu & L. Chen 2009; C.-B. Wang et al. 2013; K. Fan et al. 2021). However, this is not accurate when $B_w^*$ is extremely large, as there exists an upper threshold $B_w^{*,+}$ that increases with $k_x$, as shown in Figure 6(a). When $B_w^* > B_w^{*,+}$, CR approaches 0, and chaotic motion does not occur. According to Equation (10) and Figure 6(b), for large $B_w^*$, $P^m_{\mathrm{eff.}}$ increases, causing the condition $P^m_{\mathrm{eff.}} < 25$ to no longer hold, and thus chaotic motion does not occur. This indicates that $P_{\mathrm{eff.}}$ is more closely related to the behavior of chaos than wave amplitude alone. However, the upper threshold $B_w^{*,+}$ predicted by $P_{\mathrm{eff.}}$ in Figure 6(b) slightly deviates from the simulation results shown in Figure 6(a), suggesting future work may need to refine the current formulation of $P_{\mathrm{eff.}}$ for large-amplitude scenarios.

Chaos behavior of trapped particles in magnetic mirrors has been theoretically studied by B. V. Chirikov (1960, 1987), who also found that changes in $\mu_m$ occur within a small neighborhood around the field minimum. In a magnetotail-like field, the curvature parameter $\kappa = \sqrt{R_c^m/\rho_i^m}$ can be used to determine chaos and nonadiabatic behavior (J. Buechner & L. M. Zelenyi 1989), which is similar to $P_{\mathrm{eff.}}$ proposed in this paper. In the Earth's magnetosphere, pitch-angle scattering of ring current ions and radiation belt electrons caused by the curvature of the Earth's magnetic field is considered important (S. L. Young et al. 2008; W. W. Eshetu et al. 2021; Z. Wei et al. 2025), and has been supported by observations from several missions (M. W. Chen et al. 2019; Y. Yu et al. 2020). Finite-amplitude low-frequency AWs also induce substantial magnetic field-line curvature. Our calculations indicate that the FLC pitch-angle scattering mechanism remains effective in waves. Therefore, this mechanism may play an important role in ion heating within Alfvénic turbulent plasmas, such as those found in the solar corona and solar wind. Notable prior studies on ion stochastic heating by Alfvénic turbulence emphasize the changes in perpendicular ion energy driven by gyroscale turbulent electric fields, which consist of irrotational (electrostatic-dominated) and solenoidal (induction-dominated) components (B. D. G. Chandran et al. 2010; Q. Xia et al. 2013; I. W. Hoppock et al. 2018; Z. Johnston et al. 2025). In contrast, our study highlights the role of magnetic field-line deformations in stochastic heating, which can occur without turbulent electric fields and does not require the waves to be confined to the gyroscale. We suggest that both of these stochastic heating mechanisms may operate in the corona and solar wind plasmas. A comparative analysis of these two mechanisms remains a task that still needs to be addressed in future work.

Our core analysis is performed in the wave frame, where the electric field is absent, and the magnetic field does no work on charged particles. For an individual ion, its kinetic energy remains constant in the wave frame. However, for a large number of particles initially in a cold plasma state, stochastic scattering driven by WFLC leads to unpredictable changes in velocity, manifesting the behavior of chaotic motion. This chaos process irreversibly broadens the velocity distribution, redistributing the energy from ordered bulk kinetic energy $K^{\mathrm{WF}}$ to disordered thermal kinetic energy $T^{\mathrm{WF}}$, where

$$K^{\mathrm{WF}} = \frac{N}{2} m_i \langle \boldsymbol{v} \rangle^2,$$

$$T^{\mathrm{WF}} = \frac{N}{2} m_i (\langle \boldsymbol{v}^2 \rangle - \langle \boldsymbol{v} \rangle^2) = \frac{N}{2} m_i v^2 - K^{\mathrm{WF}},$$

the superscript WF denotes the wave frame, while $\langle \cdot \rangle$ represents averaging over all particles, and $N$ is the total number of particles. In this context, we refer to this process as "stochastic heating," where the total kinetic energy of the particles $E^{\mathrm{WF}} = T^{\mathrm{WF}} + K^{\mathrm{WF}} = \frac{N}{2} m_i v^2$ remains conserved, but the thermal component $T^{\mathrm{WF}}$ increases due to the redistribution of kinetic energy from the bulk kinetic component $K^{\mathrm{WF}}$ to the thermal kinetic component $T^{\mathrm{WF}}$. To address the net particle energization (i.e., energy transfer from waves to particles), we need to shift to the plasma frame. In this frame, both the thermal kinetic energy, $T^{\rm PF} = T^{\rm WF}$, and the bulk kinetic energy, $K^{\rm PF} = \frac{N}{2} m_i \langle \boldsymbol{v}^{\rm WF} + v_{\rm A}\hat{\boldsymbol{z}} \rangle^2$, increase from zero to a positive value. This leads to a net increase in the particles' total kinetic energy, given by $E^{\rm PF} = E^{\rm WF} + \frac{1}{2} m_i v_{\rm A}^2 + m_i v_{\rm A} \langle v_z^{\rm WF} \rangle = m_i v_{\rm A} \langle v_z^{\rm WF} \rangle + {\rm const.}$, where the superscript PF denotes the plasma frame. Consequently, in the plasma frame, the wave carries a nonzero electric field that performs positive work on the particles, which leads to a net increase in the sum of thermal and bulk kinetic energies, resulting in genuine energization. The total energy transfer from waves to particles is proportional to the change in $\langle v_z \rangle$ in the wave frame.

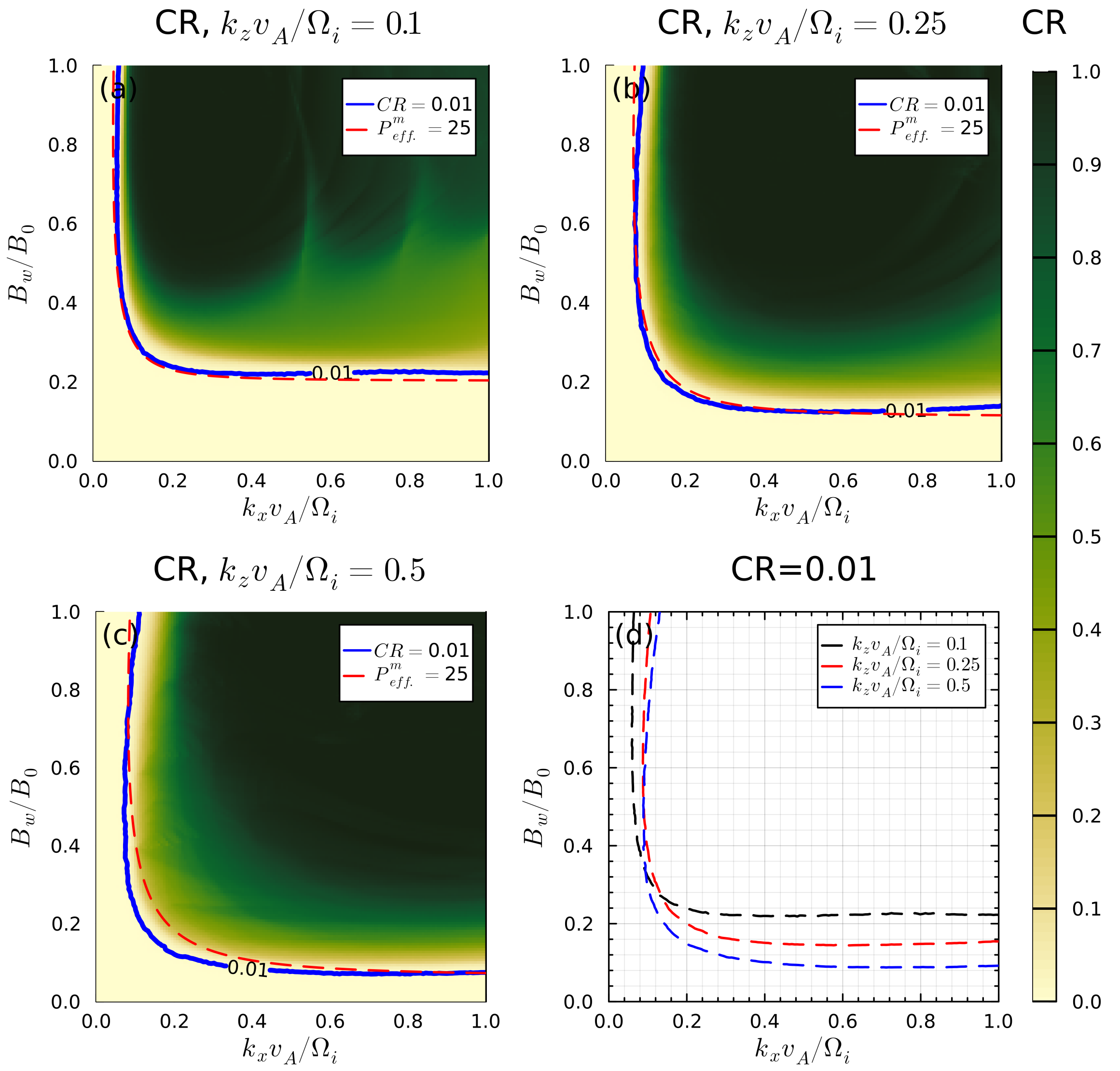

**Figure 5.** CR and contour lines of CR and $P^m_{\rm eff.}$ in $B^*_w - k^*_x$ parameter space for $k^*_z = 0.1$, 0.25, and 0.5. The CR calculation considered 4000 particles with $v = v_A$, $\theta_0$ uniformly distributed in $[0, \pi]$, $\phi_0 = 0$, and $\psi_0$ uniformly distributed in $[0, 2\pi]$. The blue lines in (a), (b), and (c) represent the CR = 0.01 contour, while the red lines correspond to the $P^m_{\rm eff} = 25$ contour. (d) CR = 0.01 contour lines for $k^*_z = 0.1$, 0.25, and 0.5.

Recent observations from the Parker Solar Probe (PSP) and Solar Orbiter (SolO) have revealed enhanced proton temperatures and strong Alfvénic fluctuations in low-frequency regimes (S. Bourouaine et al. 2024). Such conditions, characterized by large-scale magnetic field bending, may satisfy the chaos criterion $P_{\rm eff.} < 25$ derived here. Future comparisons between field-line curvature and proton pitch-angle distributions measured by PSP, SolO, or multispacecraft joint observations (Y. Zhou & C. Shen 2024; C. Shen et al. 2025) could provide direct observational tests of the proposed WFLC-induced stochastic heating mechanism.

The WFLC mechanism may also be crucial for ion heating in solar wind dissipative switchback clusters (C. Hou et al. 2023; Q. Luo et al. 2023), as these switchbacks feature large-scale magnetic field-line deformations that can generate substantial FLC. Our research contributes to enhancing and supplementing existing KAW ion heating theoretical models (M. K. Chettri et al. 2024; S. Ayaz et al. 2025a, 2025b) by incorporating nonlinear dynamics approaches. Future observational studies should aim to verify this mechanism by investigating the relationship between magnetic field-line curvature and ion pitch-angle scattering in solar wind and coronal environments.

At kinetic scales, observations from the Magnetospheric Multiscale (MMS) mission have revealed direct evidence of curvature-driven pitch-angle scattering in Earth's magnetosphere (Y. Yu et al. 2020), suggesting that similar microphysical processes operate in heliospheric plasmas. The WFLC mechanism identified here thus provides a unifying framework connecting curvature-induced nonadiabaticity observed in planetary magnetospheres to stochastic ion heating in the solar wind. Observations from the MMS spacecraft reveal a strong correlation between the ion gyroradius, $\rho_i$, resulting from KAW perpendicular heating, and the perpendicular wavelength, $\lambda_\perp$, suggesting that perpendicular heating by KAW is limited by $\lambda_\perp$ (Z. Y. Liu et al. 2023). However, this limitation can be broken when the wave amplitude exceeds the stochastic heating threshold. Therefore, future observations detecting ion distributions with $\rho_i$ larger than $\lambda_\perp$ could serve as evidence of stochastic heating events. Beyond its observational relevance, this mechanism extends existing KAW-based ion heating theories by introducing a nonlinear dynamical pathway in which macroscopic Alfvénic curvature transitions into microscopic chaos. This multiscale coupling offers new insight into how turbulent energy in the solar wind is dissipated into ion thermal energy.

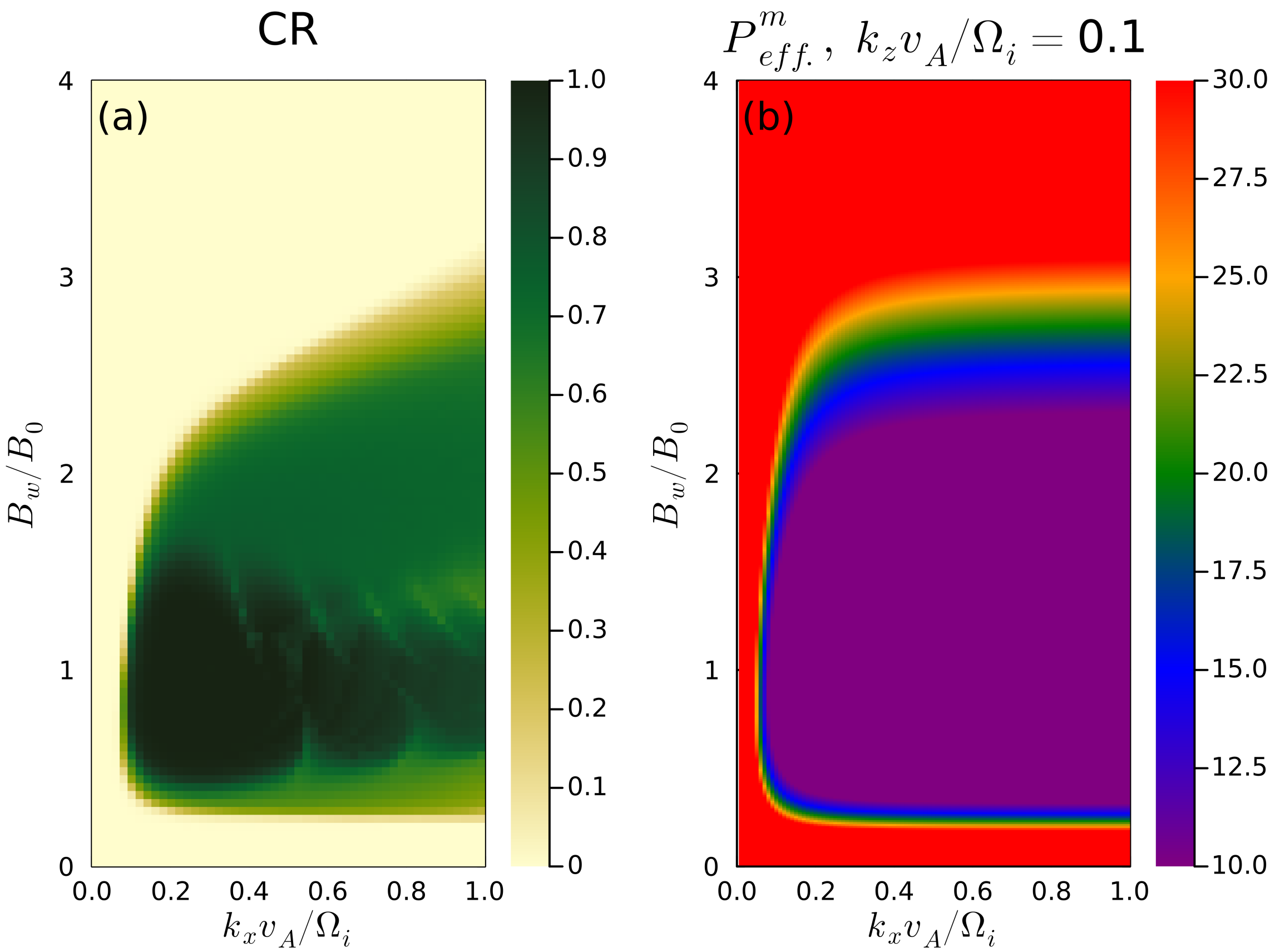

**Figure 6.** CR and $P^m_{\rm eff.}$ in $B_w^* - k_x^*$ parameter space for $k_z^* = 0.1$. The CR calculation considered 4000 particles with $v = v_A$, $\theta_0$ uniformly distributed in [0, $\pi$], $\phi_0 = 0$, and $\psi_0$ uniformly distributed in [0, $2\pi$].

## Acknowledgments

The work was supported by the National Natural Science Foundation of China (NSFC) under grants Nos. 42530105, 42241118, and 42150105; the National Key R&D Program of China under grants Nos. 2021YFA0718600 and 2022YFF0503800; the China National Space Administration (CNSA) under grants Nos. D010301, D010202, and D050103; and the Specialized Research Fund for the State Key Laboratory of Solar Activity and Space Weather. All numerical simulation data and analysis codes used in this study are openly available at J. Peng (2025a, 2025b). Numerical integrations and chaos quantification were performed using open-source Julia packages, including `DynamicalSystems.jl` (G. Datseris & U. Parlitz 2022; G. Datseris 2018), `FractalDimensions.jl` (G. Datseris et al. 2023), and `DifferentialEquations.jl` (C. Rackauckas & Q. Nie 2017), within the `Julia` computing environment (J. Bezanson et al. 2017).

## ORCID iDs

Jingyu Peng https://orcid.org/0000-0001-7537-5999
Jiansen He https://orcid.org/0000-0001-8179-417X
Rong Lin https://orcid.org/0000-0001-7655-5000